\documentclass[a4paper, amsfonts, amssymb, amsmath, reprint, showkeys, nofootinbib, twoside]{revtex4-2}
\usepackage[english]{babel}
\usepackage[utf8]{inputenc}
\usepackage[colorinlistoftodos, color=green!40, prependcaption]{todonotes}
\usepackage{amsthm}
\usepackage{mathtools}
\usepackage{physics}
\usepackage{xcolor}
\usepackage{graphicx}
\usepackage[left=23mm,right=13mm,top=35mm,columnsep=15pt]{geometry} 
\usepackage{adjustbox}
\usepackage{placeins}
\usepackage[T1]{fontenc}
\usepackage{lipsum}
\usepackage{csquotes}

\usepackage{graphicx}
\usepackage{graphics}
\usepackage{subfigure}
\usepackage{bm}
\usepackage[pdfauthor={Jeremy Sakstein}]{hyperref}
\usepackage{tabularx}
\usepackage{setspace}

\DeclareRobustCommand{\okina}{%
  \raisebox{\dimexpr\fontcharht\font`A-\height}{%
    \scalebox{0.8}{`}%
  }%
}

\graphicspath{ {./sections/images/} }

\newcommand{\msun}{{\rm M}_\odot}

\begin{document}
\title{Parameterized Post-Tolman-Oppenheimer-Volkoff Framework for Screened Modified Gravity with an Application to the Secondary Component of GW190814}

\author{Christopher Reyes}
    \email{cmreyes3@hawaii.edu}% Your name
    \affiliation{Department of Physics \& Astronomy, University of Hawai\okina i, Watanabe Hall, 2505 Correa Road, Honolulu, HI, 96822, USA}
    \author{Jeremy Sakstein} \email{sakstein@hawaii.edu}
\affiliation{Department of Physics \& Astronomy, University of Hawai\okina i, Watanabe Hall, 2505 Correa Road, Honolulu, HI, 96822, USA}

\date{\today} % Leave empty to omit a date

\begin{abstract}
The secondary component of GW190814 has  mass in the range $2.5$--$2.67{\rm  M}_\odot$, placing it within the lower mass gap separating neutron stars from black holes.~According to the predictions of general relativity and state-of-the-art nuclear equations of state, this object is too heavy to be a neutron star.~In this work, we explore the possibility that this object is a neutron star under the hypothesis that general relativity is modified to include screening mechanisms, and that the neutron star formed in an unscreened environment.~We introduce a set of parameterized-post-Tolman-Oppenheimer-Volkoff (post-TOV) equations appropriate for screened modified gravity whose free parameters are environment-dependent.~We find that it is possible that the GW190814 secondary could be a neutron star that formed in an unscreened environment for a range of reasonable post-TOV parameters.

\end{abstract}

\maketitle

\section{Background and motivation} \label{sec:outline}

Gravitational wave (GW) observations  of compact objects such as black holes (BHs) and neutron stars (NSs) made by the LIGO/Virgo/KAGRA interferometers are providing us with a deeper understanding of these objects, and the laws of physics that govern them.~The majority of events observed to date are consistent with the predictions of general relativity, stellar structure theory, and nuclear physics;~but there are a small number of events that cannot be accommodated within these paradigms.~This paper is concerned with one such event --- GW190814 \cite{LIGOScientific:2020zkf}.~The progenitor system for this event is the binary merger of an approximately $23{\rm M}_\odot$ BH with a compact object of undetermined nature with mass in the range $2.5\msun\le M\le2.67\msun$.~This secondary object lies within the lower mass gap that separates black holes from neutron stars.~There are no stellar evolution pathways to form a black hole this light but it has been suggested that this object could be a primordial black hole \cite{Carr:2019kxo,Clesse:2020ghq}, although it has been claimed that such a scenario is unlikely \cite{Vattis:2020iuz}.~If instead the secondary compact object is a neutron star then it is heavier than the maximum neutron star mass implied by general relativity (GR) applied to realistic nuclear equations of state.~If this object is a neutron star then it can be explained by non-standard/exotic equations of state e.g., \cite{Bombaci:2020vgw,Godzieba:2020tjn,Thakur:2022qpy,Kayanikhoo:2023yie,Zuraiq:2023bpw,Goswami:2023tps};~rapid rotation e.g., \cite{Most:2020bba,Zhang:2020zsc,Biswas:2020xna,Li:2020ias,Zhou:2020xan,Ghosh:2022qkq};~magnetic fields \cite{Rather:2021azv,Kayanikhoo:2023yie,Zuraiq:2023bpw};~dark matter/new particles \cite{Lee:2021yyn,Das:2021yny,Guha:2021njn,Sen:2022pfr};~or modified gravity \cite{Astashenok:2020qds,Nunes:2020cuz,Moffat:2020jic,Astashenok:2021xpm,Astashenok:2021peo,Barros:2021jbt}.~In this paper, we explore the possibility that the secondary object in GW190814 is a neutron star that formed under the influence of screened modified gravity.~

Screened modified gravity refers to extensions of GR that include \textit{screening mechanisms} \cite{Joyce:2014kja,Burrage:2016bwy,Burrage:2017qrf,Sakstein:2018fwz,Baker:2019gxo,Brax:2021wcv}.~These theories introduce new light degrees of freedom --- typically scalars --- that couple to matter.~Modifications of GR such as these typically run afoul of solar system tests of gravity or laboratory searches for fifth-forces (depending on the mass of the new particle) \cite{Adelberger:2003zx,Kapner:2006si,Will:2014kxa,Safronova:2017xyt,Lee:2021yyn} but screening mechanisms evade this via non-linearities in the equations of motion.~This feature has led to them becoming a fundamental building block of dark energy theories that utilize light scalars to drive the present phase of cosmic acceleration.~Examples of screening mechanisms include the chameleon \cite{Khoury:2003aq,Khoury:2003rn,Brax:2004qh}, symmetron \cite{Hinterbichler:2010es,Hinterbichler:2011ca}, dilaton \cite{Brax:2010gi}, and Vainshtein \cite{Vainshtein:1972sx,Nicolis:2008in} mechanisms, as well as dark matter-baryon interactions \cite{Sakstein:2019qgn} and K-mouflage \cite{Babichev:2009ee,Brax:2012jr} gravity.

The fundamental field theory descriptions of these mechanisms are diverse, but all share one common feature:~the gravitational field sourced by matter is environment-dependent.~Screened dark energy theories typically have deviations from GR becoming stronger in environments that are less dense than the solar system.~This ensures that modified gravity is able to simultaneously drive the Universe's acceleration --- a low density cosmological phenomenon --- while evading fifth-force constraints in the solar system.~In this work, we utilize this generic environmental-dependence to derive a parameterized post-Tolman-Oppenheimer-Volkoff (post-TOV) framework for describing neutron stars under screened modified gravity in a theory-independent manner.~The free parameters of our framework coincide with some parameters appearing in the parameterized post-Newtonian (PPN) framework for testing gravity in the solar system.~The effects of screening mechanisms are incorporated by allowing these parameters to be environment-dependent so that PPN bounds do not apply.~We solve the post-TOV equations for a range of parameters and find neutron star masses compatible with the secondary object in GW190814.~It is therefore possible that this object is a neutron star that formed in an unscreened environment.~We discuss follow-up work that could confirm this scenario in our conclusions.

This work is organized as follows.~In section \ref{sec:post-TOV} we introduce our post-TOV equations for screened modified gravity.~In section \ref{sec:results} we solve these equations for realistic equations of state.~We discuss our results and conclude in section \ref{sec:conclusions}.

\section{Parameterized Post-Tolman Oppenheimer-Volkoff Equation} 
\label{sec:post-TOV}

Our starting point for deriving the post-TOV framework for screened modified gravity is the post-TOV formalism derived by \cite{Glampedakis:2015sua} (see  \cite{1983ApJ...275..867C} for earlier work).~The framework is a parameterization of the TOV equation up to second post-Newtonian (2PN) order: 
\begin{align}
\label{eq:postTOVp}
    \frac{dp}{dr} &= -\frac{G(\varepsilon + p)(m+4\pi r^{3}p)}{1-2Gm/r} -\frac{G\rho m}{r^2}(P_{1} +P_{2})\\
    \frac{dm}{dr} &= 4\pi r^{2}\varepsilon +4\pi r^2\rho( M_{1}+M_{2})
    \label{eq:postTOVm}
\end{align}
where 
\begin{align}
P_{1} &= \delta_{1}\frac{Gm}{r} +4\pi \delta_{2}\frac{r^3 p}{m}\\
M_{1} &= \delta_{3}\frac{Gm}{r} +\delta_{4}\Pi\\
P_{2} &= \pi_{1}\frac{G^2m^3}{r^5\rho} +\pi_{2} \frac{G^2m^2}{r^2} + \pi_{3}Gr^{2}p + \pi_{4}\Pi\frac{p}{\rho}\\
M_{2} &= \mu_{1}\frac{G^2m^3}{r^5\rho} +\mu_{2} \frac{G^2m^2}{r^2} + \mu_{3}Gr^{2}p + \mu_{4}\Pi\frac{p}{\rho} \\\nonumber&+\mu_{5}\Pi^{3}\frac{r}{Gm}.
\end{align}
In the equations above, $\varepsilon$ is the energy density, $p$ is the pressure, and $\rho$ is the rest mass density equal to $m_{n}n_{b}$ with $m_{n}$ is the mass of the neutron and $n_{b}$ the baryon number density, and $\Pi  = (\varepsilon-\rho)/\rho$ is the internal energy per unit mass.~The constants $\delta_{i}$, $\pi_{i}$, and $\mu_{i}$ are free parameters quantifying the strength of deviations from GR.~These parameters are zero in GR.~The coefficients $\delta_{i}$ are linear combinations of PPN parameters, and the terms they multiply are the 1PN post-TOV corrections.~The PPN coefficients are well constrained in the Solar system, $\lvert \delta_{i} \rvert \ll 1$ \cite{Will:2014kxa}, implying that 1PN corrections to the TOV equations are negligible.~This led reference \cite{Glampedakis:2015sua} to introduce the terms proportional to $\pi_{i}$ and $\mu_i$, which are 2PN and not constrained in the solar system.~

In theories that include screening mechanisms, the parameters $\delta_{i}$, $\pi_{i}$, and $\mu_{i}$ are environment-dependent --- they are set by the screening level of the galaxy where the neutron star is located.~Since screened modified gravity theories are not subject to solar system tests\footnote{Strictly speaking, solar system tests do impose bounds e.g., \cite{Sakstein:2017pqi} but these are far weaker than the equivalent bounds on theories that do not include screening mechanisms.}, the 1PN corrections in unscreened environments are the leading-order corrections to GR, so we set  $\pi_{i}=0$ and $\mu_{i}=0$ from hereon.~The remaining post-TOV parameters are related to the standard PPN parameters as follows (see \cite{Will:2014kxa,Ip:2015qsa} for the definitions of these):
\begin{align}
\delta_{1} &= 3(1+\gamma) -6\beta+\zeta_{2},\label{eq:delta1}\\
\delta_{2} &= \gamma-1  + \zeta_{4},\label{eq:delta2}\\
\delta_{3} &= \frac{1}{2}(12\beta-\gamma-11-\zeta_{2} + 2\zeta_{4}),\label{eq:delta3}\\
\delta_{4} &=\zeta_{3}.\label{eq:delta4}\\
\end{align} 
We can simplify these further by noting that the parameters $\zeta_i$ are only non-zero in theories that do not conserve energy and momentum i.e., those that cannot be derived from diffeomorphism-invariant actions \cite{Will:1972zz,Lee:1974nq}.~All known screening mechanisms are derived from field theory actions, so we set $\zeta_i=0$ without loss of generality.~In addition to the free parameters $\gamma$ and $\beta$, we introduce a third parameter $\omega$ defined via $G = G_{N}(1+\omega)$ where $G_N$ is the value of Newton's constant measured in the laboratory.~This parameter accounts for violations of the strong equivalence principle predicted by screened modified gravity theories whereby the value of $G$ differs between weakly and strongly gravitating objects \cite{Sakstein:2017pqi}.~Our post-TOV formalism for screened modified gravity is therefore described by three free parameters --- $\gamma$, $\beta$, and $\omega$.~In GR one has $\gamma=1$, $\beta=1$, and $\omega=0$.~Previous work \cite{Nunes:2020cuz} has investigated whether chameleon gravity can account for the secondary object in GW190814 as a neutron star.~In that work, only violations of the strong equivalence principle were considered i.e., the effects of $\omega$.~Our post-TOV formalism generalizes this to include $\gamma$ and $\beta$, and to cover a larger class of screening mechanisms.

\begin{figure*}
     \centering
%     \begin{subfigure}
     \centering
     \includegraphics[width=1\linewidth]{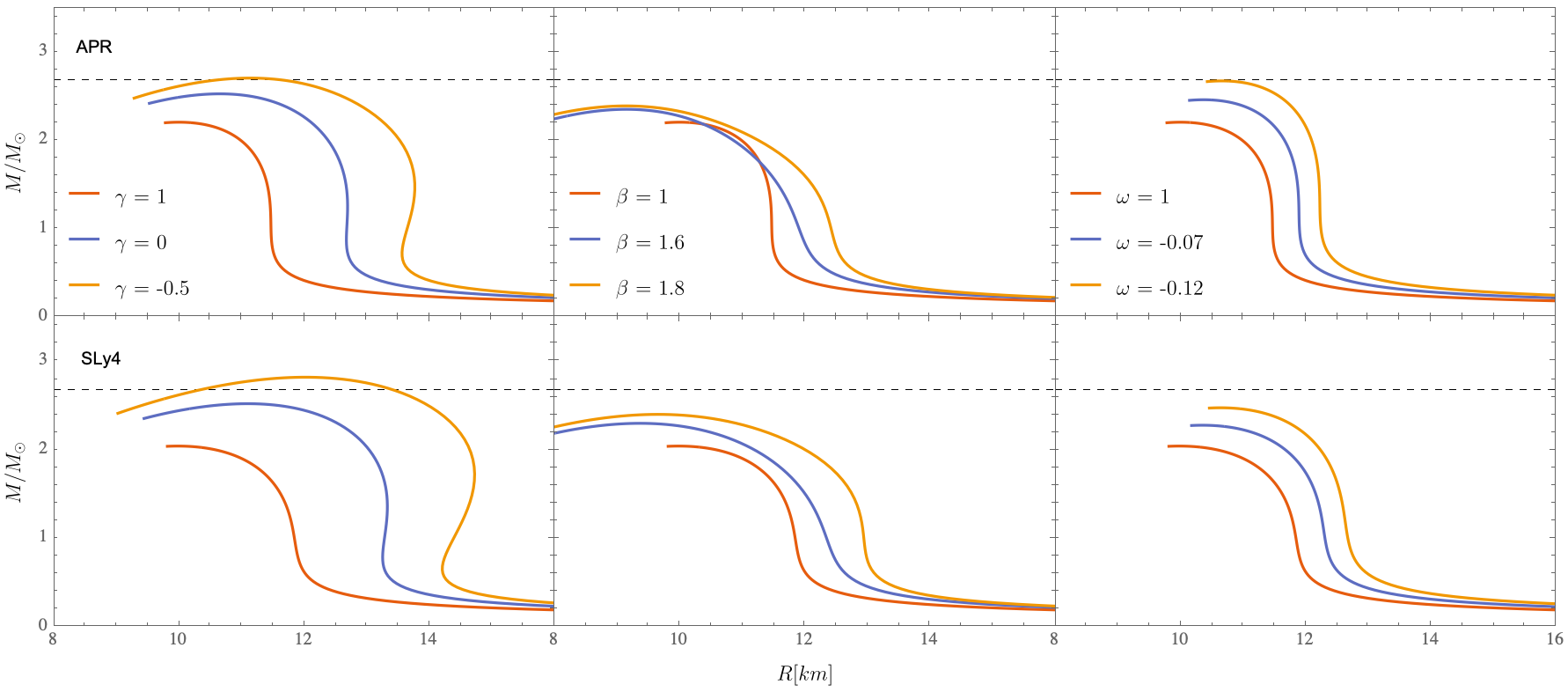}
     \caption{Mass-radius relations when the post-TOV parameters are varied individually with all others fixed to their GR values. The upper panels were computed using the APR EOS and the lower panels were computed using the SLY4 EOS. The left panels show the effect of varying $\gamma$, the center panels show the effect of varying $\beta$, and the right panels show the effect of varying $\omega$. The parameters corresponding to each curve are given in the panels. The GR values of the post-TOV parameters are $\gamma=\beta=1$, $\omega=0$. }     %\end{subfigure}
     \label{fig:Mass/Radius}
 \end{figure*}

Before proceeding to solve our post-TOV equations, we delineate the conditions under which they apply, and the screening mechanisms to which they appertain.~The PPN formalism makes the assumption that there are no new mass scales in the theory so that the expansion is in inverse powers of $r$ i.e., there are no Yukawa-like terms.~This assumption is exact for screened dark matter-baryon interactions and chameleon, symmetron, and dilaton theories with parameters such that the scalar's (environment-dependent) Compton wavelength (inverse-mass) in the neutron star's environment is larger than $\sim$ $15$ km (the largest typical neutron star radius).~The latter condition is expected to hold in screened dark energy theories because the scalar's Compton wavelength is, by design, larger than $\mathcal{O}(100{\rm kpc})$.~Theories that exhibit Vainshtein and K-mouflage screening do not fit into the PPN framework because they require additional terms associated with the high-degree of non-linearity in these theories \cite{Avilez-Lopez:2015dja,Bolis:2018kcq,Renevey:2020tvr}.

\section{Results} \label{sec:results}

We solved the post-TOV equations \eqref{eq:postTOVp} and \eqref{eq:postTOVm}  numerically for reasonable values of the parameters $\gamma$, $\beta$, and $\omega$ to derive the neutron star mass-radius relation using the APR and SLY4 equations of state.~These are realistic equations of state that are commonly employed in the literature but, assuming GR, they are excluded by observations of GW170817 \cite{Radice:2017lry}.~They are not excluded in screened modified gravity because GW170817 has yet to be modeled in the same level of detail in these theories --- numerical waveforms have not been computed and magnetohydrodynamic simulations have not been performed.~Figure~\ref{fig:Mass/Radius} shows the mass-radius relation when each post-TOV parameter is varied individually with all others fixed to their GR values.~The effects of each parameter is seen to be the following.
\begin{itemize}
    \item[$\bm\omega$:] Negative values of $\omega$ increase the maximum neutron star mass.~This can be understood as follows.~Negative $\omega$ implies that gravity is weaker in unscreened environments than in the solar system.~Scaling out dimensionful quantities from the (post-)TOV equations shows that the maximum NS mass scales as $M\propto G^{-3/2}$ (see e.g., \cite{Vijaykumar:2020nzc}).~The weakening of gravity therefore increases the maximum NS mass.~Our results for varying $\omega$ are consistent with those of \cite{Nunes:2020cuz}.
    \item[$\bm\gamma$:] Reducing the value of $\gamma$ below the GR value of unity increases the maximum NS mass.~This can be understood by considering the definitions of the $\delta_i$ parameters defined in equations~\eqref{eq:delta1}--\eqref{eq:delta4}.~For $\gamma<1$ and $\beta$ fixed to its GR value of unity, $\delta_1$ and $\delta_2$ are negative and $\delta_3$ is positive.~The terms that $\delta_1$ and  $\delta_2$ multiply in \eqref{eq:postTOVp} are positive, making the post-TOV contribution to $\dd p/\dd r$ positive.~The pure GR contribution to the TOV equation is negative, so the effects of $\delta_1$ and $\delta_2$ are tantamount to weakening gravity, which, as above, implies heavier NSs.~The effect of $\delta_3$ is to add a positive contribution to equation \eqref{eq:postTOVm}.~One would then expect heavier stars at fixed radius because the effective energy density sourcing the mass is increased.~This effect is evident in figure~\ref{fig:Mass/Radius}.
    
    \item[$\bm \beta$:] Increasing $\beta$ has the effect of increasing the maximum NS mass.~Examining equations ~\eqref{eq:delta1}--\eqref{eq:delta4}, it is evident that for $\gamma=1$ (the GR value), increasing $\beta$ above its GR value of unity implies that $\delta_1$ is negative.~Similar to increasing $\gamma$, this implies that the contribution of the terms they multiply in equation \eqref{eq:postTOVp} behaves as if gravity is weakened, implying a larger maximum NS mass.~$\delta_3$ becomes increasingly positive with increasing $\beta$.~This adds an extra effective energy density to equation~\eqref{eq:postTOVm}, resulting in heavier stars at fixed radius as evidenced in figure~\ref{fig:Mass/Radius}.~
\end{itemize}

\begin{figure}
    \centering
     \includegraphics[width=1\linewidth]{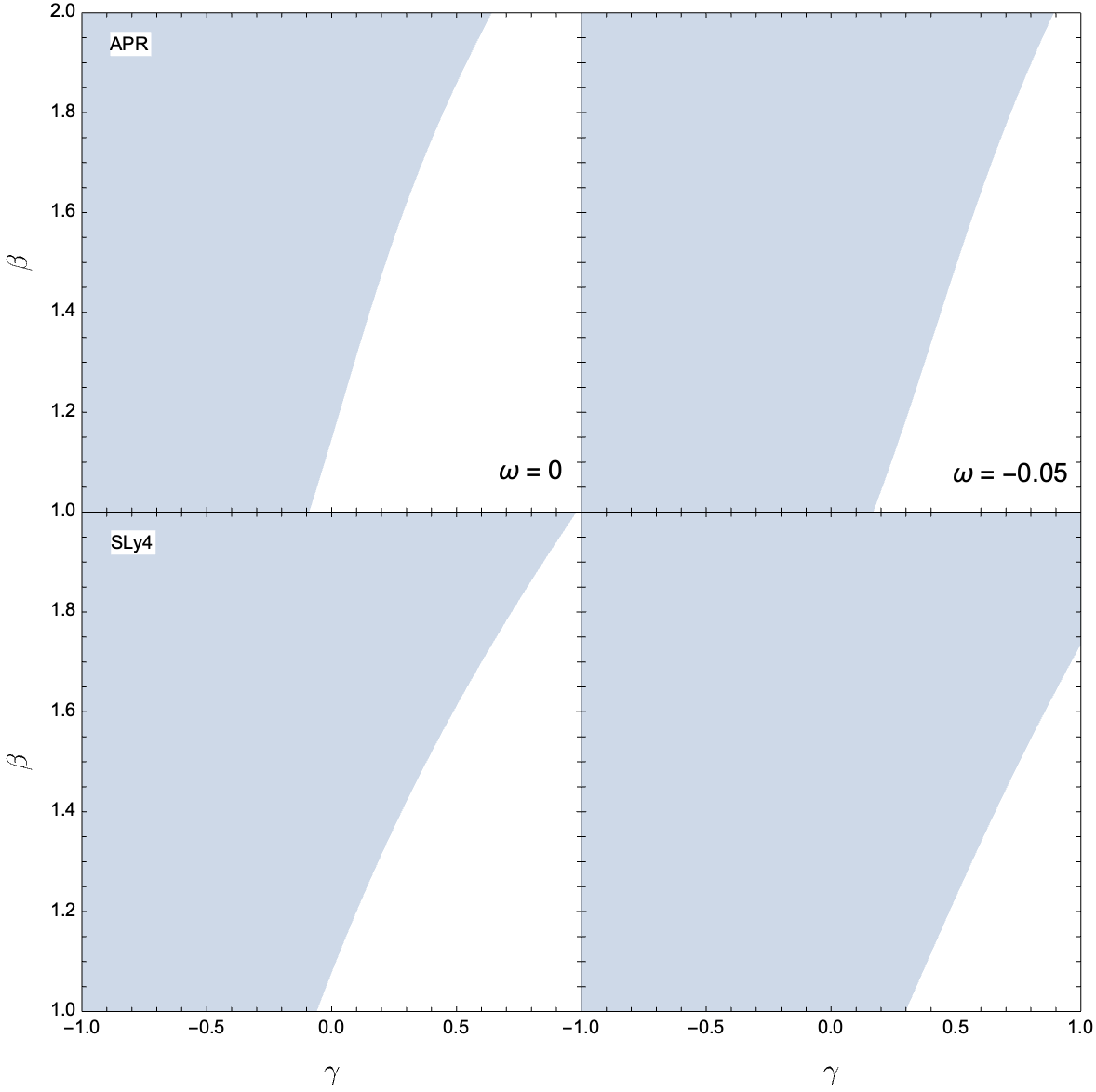}
    \caption{Values of $\gamma$ and $\beta$ (blue shaded region) where screened modified gravity can explain the secondary component of GW190814 as a neutron star.~The left panels correspond to $\omega=0$ and the right to $\omega=-0.05$.~The top panels were computed with the APR EOS and the bottom with the SLY4 EOS.}
    \label{fig:region}
\end{figure}

In practice, one expects all three post-TOV parameters to simultaneously deviate from their GR values in screened modified gravity, so in figure~\ref{fig:region} we show the regions of post-TOV parameter space where the maximum NS mass is compatible with the mass of the secondary object in GW190817.~Evidently, there are large regions of parameter space where screened MG can accommodate the secondary object as a NS.~Typical theories have $\gamma$ and $\beta$ varying from their the GR values by order unity \cite{Sakstein:2017pqi}, and astrophysical bounds typically constrain variations in $G$ to be smaller than $\sim5\%$ i.e., $|\omega|\le0.05$ \cite{Jain:2012tn,Desmond:2019ygn,Desmond:2020wep}, so the parameters where GW190814 is accounted for are realistic.

\section{Discussion and Conclusions} \label{sec:conclusions}

Our results ostensibly indicate that if GR is modified to include screening mechanisms, the secondary object in GW190814 could be a neutron star that formed in an unscreened environment.~However, before concluding that this is indeed the case, there are caveats that must be considered.

The first caveat is that the post-TOV parameters compatible with GW190814 may not be compatible with theory-specific bounds from other probes.~In the case of dark matter--baryon screening, there are no bounds on $\omega<0$  nor on $\gamma$ and $\beta$ \cite{Desmond:2020wep}.~The situation for chameleon, symmetron, and dilaton theories is more complex.~There are a plethora of bounds from a diverse array of probes \cite{Burrage:2016bwy,Burrage:2017qrf,Sakstein:2018fwz,Baker:2019gxo,Brax:2021wcv}, but these are difficult to translate into our post-TOV parameters $\omega$, $\gamma$, and $\beta$.~Generically, the values of these parameters shown in figure~\ref{fig:region} are reasonable provided that both the host galaxy and the neutron star are unscreened.~This is unlikely for the case of Hu \& Sawicki $f(R)$ chameleon gravity \cite{Hu:2007nk} --- the quintessential paradigm for chameleon screening on astrophysical scales --- because astrophysical bounds imply that nearly all galaxies will be screened \cite{Desmond:2020gzn}.~The screening level of more general chameleon models is less definitive, as is the screening of galaxies in symmetron and dilaton theories \cite{Burrage:2023eol}.~

The second caveat is that we have neglected two-body effects on screening when drawing conclusions.~This is justified because the screening mechanisms parametrized by our post-TOV formalism are subject to no-hair theorems.~Chameleons, symmetrons, and dilatons are conformal scalar-tensor theories, which do not give rise to black hole scalar hair \cite{Bekenstein:1971hc,Sotiriou:2015pka}.~Similarly, dark matter-baryon screening mechanisms fall into the class of theories covered by the no-hair theorem derived by reference \cite{Hui:2012qt}.

The final caveat is that we are assuming that the LIGO/Virgo/KAGRA mass measurement --- which assumes that GR describes the merger process --- is robust to modified gravity effects.~It is presently unknown how screened modified gravity effects the dynamics of NS-BH mergers, but one might speculate that there is some amount of dipolar radiation owing to the need to radiate the NS's scalar charge before the formation of the final BH, which, as discussed above, cannot support scalar hair.~Quantitative investigations of NS-BH mergers in modified gravity theories are a challenging and computationally intensive task, but they may be justified if follow-up investigations of the scenario we have explored in this work provide additional validation.

Based on the considerations above, we conclude that it is possible that the secondary object in GW190814 could be a neutron star that formed under the influence of screened modified gravity, and this possibility merits further investigations.~The next logical step would be to extend our post-TOV formalism to second-order in slow rotation.~This would enable computations of the I-Love-Q relations in a theory-independent manner.~Tidal distortions were not detected in the GW190814 signal --- hence the difficulty in determining the secondary's nature as a BH or NS --- so such a computation would not enable additional tests of the screened modified gravity hypothesis from this event.~It would however provide additional constraining power from future GW190814-like events where the tidal distortions are measured.~Similarly, constraints could be obtained for NS-NS and BH-NS mergers presently observed by LIGO/Virgo/KAGRA.~Another line of inquiry would be to simulate the binary star progenitor system for this event under screened modified gravity, although this would need to be done on a theory-by-theory basis.~

\section*{Acknowledgements} \label{sec:acknowledgements}
    We are grateful for discussions with Kazuya Koyama, Hector O.~Silva, and  Nico Yunes.

\bibliography{sections/cit}

\end{document}